\documentclass[12pt]{article}

\usepackage{amsmath,amssymb,amsfonts,graphicx}
\usepackage{ucs}
\usepackage{epstopdf}
\usepackage{color}
\usepackage[utf8]{inputenc}
\usepackage{amsthm}
\usepackage{fontenc}
\usepackage{tikz}
\usepackage[dvips]{hyperref}
\RequirePackage{slashed}
\usepackage{cancel}
\usepackage[normalem]{ulem}

\usepackage{color}
\usepackage{latexsym}
\usepackage{amssymb}
\usepackage{booktabs}
\usepackage{comment}
\usepackage{authblk}

\usepackage[a4paper,top=2cm,bottom=2cm,left=2cm,right=2cm]{geometry}

\newcommand{\bit}{\begin{itemize}}
\newcommand{\eit}{\end{itemize}}

\def\be{\begin{equation}}
\def\ee{\end{equation}}
\def\bea{\begin{eqnarray}}
\def\eea{\end{eqnarray}}

\newlength\savedwidth

\title{ A { model calculation of}  double parton distribution functions of 
the pion }

\author[1]{Matteo Rinaldi}
\affil[1]{\small{ Departament de Fisica Te\`orica, Universitat de Val\`encia
and Institut de Fisica Corpuscular, Consejo Superior de Investigaciones
Cient\'{\i}ficas, 46100 Burjassot (Val\`encia), Spain } }
\author[2]{Sergio Scopetta}
\affil[2]{ \small{ Universit\`a degli Studi di Perugia, and Istituto Nazionale 
di Fisica 
Nucleare,
Sezione di Perugia. Perugia  Via A. Pascoli, I-06123,Italy } }

\author[3]{Marco Traini}
\affil[3]{ \small{ INFN - TIFPA, Dipartimento di Fisica, Universit\`a degli 
Studi 
di Trento, Via Sommarive 14, I-38123 Povo (Trento), Italy } }
\author[1]{Vicente Vento}

\begin{document}

\maketitle

\begin{abstract}
{
Two-parton correlations { in the pion}  are
investigated  in terms of double parton 
distribution functions.  
A Poincar\'e covariant 
Light-Front framework has been adopted. 
As non perturbative input, the 
pion wave function obtained within the so-called 
soft-wall AdS/QCD model has been used.
Results show how novel dynamical information 
on the structure of the pion,  {  not accessible 
through one-body parton distribution,} are encoded in double
parton distribution functions.  }
\end{abstract}

\section{\label{sec:intro}Introduction}

Double parton scattering (DPS), the 
simplest form of multiple parton interaction (MPI), has been 
{ observed at the} LHC (see, e.g., Ref. \cite{Aad:2013bjm}).
The DPS cross section can be written in terms of double 
parton distribution functions (dPDFs) \cite{Paver:1982yp,Diehl1},
which represent the number density of two partons located at a given 
transverse separation in 
coordinate space and with given longitudinal momentum fractions. 
This is an information complementary to the tomography accessed through
electromagnetic probes 
in terms of generalized parton distributions (GPDs)
\cite{Guidal:2013rya,Dupre:2016mai}.  
If measured, dPDFs would therefore represent 
a novel tool to access the three-dimensional hadron structure.
However, since dPDFs describe soft
Physics, they are non perturbative objects and have
not been evaluated in 
QCD. It is therefore useful to
estimate them at low momentum scales ($\sim \Lambda_{QCD}$),
for example using quark models as has been proposed in Refs. 
\cite{bag,noi2,noi1,plb,JHEP2016,Traini:2016jru}. 
In order to match theoretical predictions with {future   } experimental 
analyses, 
the results of these 
calculations are then evolved using perturbative QCD to reach the 
{ high momentum} scale 
of the data \cite{Kirschner:1979im,Shelest:1982dg}.

In {  a }  previous work, {  use has been 
made of }  the AdS/QCD framework to study dPDFs in 
proton-proton collisions  \cite{Traini:2016jru} . The  
AdS/QCD approach establishes a 
correspondence between conformal field theories and gravitation in an 
anti-de-Sitter space \cite{Maldacena:1997re,Witten:1998qj}. The so-called 
bottom-up approach implements important features of QCD, generating a theory 
in which 
conformal symmetry is  { restored} asymptotically  
\cite{Polchinski:2000uf,Brodsky:2003px,Erlich:2005qh,DaRold:2005mxj}. This 
approach has been successfully applied to the description of the spectrum of 
hadrons, { of their} form factors { (ffs)} and parton distributions 
{ (PDFs)}~ 
\cite{Br1,Br2,rho,universal,Traini:2016jko,rinaldiGPD}.
 {  In particular } the structure of the pion is an 
 interesting 
subject which has 
  attracted much 
attention from the point of view of AdS/QCD~\cite{Br1,15sp,pasquini1}.
{  In this scenario }
we 
proceed  here to generalize   { the} formalism
{ developed for nucleon dPDFs} to mesons 
and apply it to pion wave functions defined via the AdS/QCD correspondence. 
This analysis has been partially 
motivated by 
 { a}    first estimate 
of 
moments of quantities related to pion dPDFs in the lattice, { recently}   
reported \cite{Zimmermann:2017ctb}. 

In section \ref{II} we 
describe the meson dPDF
in terms of the light-front (LF) wave function (w.f.),
and introduce {  an } approximation which relates dPDFs 
to 
 GPDs and 
 ffs. Furthermore we 
  { introduce}  { a quantity relevant to DPS phenomenology}, 
the  
effective cross section $\sigma_{eff}$, in 
terms of dPDFs and  PDFs. 
In section \ref{III}   AdS/QCD model calculations of 
dPDFs are    { summarized}  and their properties analyzed.
In sec. \ref{IV} the evolution of the dPDFs to high momentum scale  
is calculated and its implications discussed.
Conclusions are collected in sec. \ref{VI}.

\section{\label{II}Double PDF and the meson light-front wave function}

In this section we describe  { how } to express
dPDFs  in terms of the  { LF } meson wave function.
{ The  formalism { we use} has been { also}  presented} in 
Ref. \cite{kase}, where 
dPDFs have been studied for a  dressed quark target treated as { a} 
two body system. 
Formally, dPDFs are defined by means of the
light-cone correlator \cite{Diehl1},

\begin{align}
\label{f2c1}
 & f'_2(x_1,x_2, {\bf k_\perp} ) = {P^+ \over 4 } \int d^2 {\bf y_\perp}
  e^{-i 
{\bf 
y_\perp} \cdot {\bf k_\perp} }\int dy^-
 \int dz^-_1dz_2^-
 \\
\nonumber
&\times
 {e^{-i x_1P^+ z^-_1-i 
x_2P^+ z^-_2} \over (2\pi)^2 }
  \langle A, {\bf 0}| \mathcal{O}(0,z_1) 
\mathcal{O}(y,z_2)|A, {\bf 
 {\bf 0}} \rangle \Big|_{y^+ =z^+_1=z_2^+=0}^{{\bf z_{1\perp}= z_{2\perp}  }=0  
}~,
\end{align}
where, for generic 4-vectors $y$ and $z$, the operator $\mathcal{O}(y,z)$ 
reads:
 
 \begin{align}
 \label{op}
  \mathcal{O}(y,z) = \bar q\left( y- {1 \over 2} z  \right) \gamma^+  q \left(  
 y+ {1 \over 2}z \right)~,
 \end{align}
\\  and $q(z)$ is the LF quark field operator.
In order to find a suitable expression of the dPDF, we  { 
make use of} the  LF wave 
function 
representation approach~
\cite{48j1,24p}. In particular, {by taking into 
account only  } the ``valence'' 
contribution in the { LF} intrinsic frame, {
the pion state is written as}

\begin{align}
\label{state}
\nonumber
 |A, {\bf P_\perp }  \rangle &= \sum_{h, \bar h} \int 
\dfrac{dx_1~dx_2}{\sqrt{x_1 x_2}} \dfrac{d^2 {\bf k_{1\perp}} d^2 \bf k_{2 
\perp} }{2(2 \pi)^3}\delta^{(2)}({\bf k_{1 \perp}+ k_{2\perp}} )
\\
\nonumber
&\times|x_1, {\bf k_{1 \perp}}+ x_1 {\bf P_\perp},h \rangle |x_2, 
{\bf k_{2 
\perp}}+ x_2 {\bf P_\perp},\bar h \rangle
\\
&\times ~\delta(1-x_1-x_2) \psi_{h, 
\bar h} (x_1,x_2,{\bf k_{1 \perp},k_{2\perp}}  )~.
\end{align}
 \\
 { Here,}
$h$ and $\bar h$  { represent }
parton helicities, $x_i = k_i^+/P^+$ and ${\bf k_{i 
\perp}}$ the 
quark longitudinal momentum fraction and its transverse momentum, respectively, 
$P^\mu$  the meson 4-momentum. The light cone components are defined 
 {  by}
$l^\pm = l^0 \pm l^3$. 
In Eq. (\ref{state}),  $ \psi_{h, 
\bar h} (x_1,x_2,{\bf k_{1 \perp},k_{2\perp}}  )$ is the  LF meson 
wave-function, whose normalization  { is } chosen as
\begin{align}
\label{norm}
 1&= {1 \over 2} \sum_{h,\bar h} \int dx_1 dx_2  {d^2 {\bf k_{1\perp} }   d^2 
{\bf k_{1\perp} }\over 16 \pi^3}  \delta(1-x_1-x_2)
\\
\nonumber
&\times \delta^{(2)}({\bf k_{1 \perp} {\bf +} k_{2\perp}}) |\psi_{h, 
\bar h} (x_1,x_2,{\bf k_{1 \perp},k_{2\perp}}  )  |^2 ~.
\end{align}
  { The w.f.  $ \psi_{h, 
\bar h} (x_1,x_2,{\bf k_{1 \perp},k_{2\perp}}  ) $   } determines the 
structure of the state   {  and is not known}.  

{
However    { one} can  obtain the dPDF by using a standard 
procedure (see e.g. Ref. 
\cite{noi1} for the proton ) which makes use of the quark-antiquark  field 
operator \cite{Br1}, the definition of the meson state Eq. (\ref{state}), of 
Eq. (\ref{f2c1}) and the anticommutation relations between 
creation-annihilation operators (see Ref. \cite{Br1} for details). The result 
of the calculation for the pion dPDF is 
}

\begin{align}
\nonumber
 f'_2(x_1,x_2, {\bf k_\perp} ) &= {1 \over 2} \sum_{h,\bar h} \int {d^2{\bf 
k_{1\perp}} 
\over 2 (2 
\pi)^3}  \psi_{h, 
\bar h} (x_1,x_2,{\bf k_{1 \perp}},-{\bf k_{1 \perp}}   )
\\
\nonumber
&\times\psi_{h, 
\bar h}^{*} (x_1,x_2,{\bf k_{1 \perp}+k_\perp},-{\bf k_{1 \perp}-k_\perp}  )
\\
&\times \delta(1-x_1-x_2)
\label{f2m1}
\\
&= f_2(x_1,{\bf k_\perp} )\delta(1-x_1-x_2).  
\label{fdiv}
\end{align}
The physical object { of interest here}  is $f_2(x_1, {\bf k_\perp} )$,  
obtained as integral over 
$x_2$ of $f'_2(x_1,x_2, {\bf k_\perp})$  and given by

\begin{align}
\label{f2m}
 f_2(x, {\bf k_\perp} ) = {1 \over 2} \sum_{h,\bar h} 
 \int {d^2{\bf k_{1\perp}} 
\over 2 (2 
\pi)^3}  \psi_{h, 
\bar h} (x,{\bf k_{1 \perp}}  ) \psi_{h, 
\bar h}^{*} (x,{\bf k_{1 \perp}+k_\perp}  ).  
\end{align}
Notice that for ${\bf k_\perp}=0 $, the usual { LF} PDF expression is 
recovered  \cite{pasquini1}. 
We will calculate the quantity $f_2(x, {\bf k_\perp} )$ 
{ encoding} the relevant dynamical information.

Since the  LF meson wave function is evaluated  { under } 
the conditions $x_2 = 1-x_1$ and $ {\bf k_{2 \perp}}=-{\bf k_{1 \perp} }   $, 
  {  due to momentum conservation, }  { for simplicity, }   we use 
the notation

\begin{align}
 \psi_{h, 
\bar h} (x_1,{\bf k_{1 \perp}}  )= \psi_{h, 
\bar h} (x_1,1-x_1,{\bf k_{1 \perp}},{\bf -k_{1\perp} }  )~.
\end{align}
\\
  We 
are mainly interested in  non 
perturbative aspects of the 
dPDFs,  { so that,} in order to   { emphasize } the role of correlations 
between $x$ and ${\bf 
k_\perp}$, { in the next sections}
the following ratio will be  
calculated:

\begin{align}
\label{rk}
 r_k(x, k_\perp) = { f_2(x,k_\perp)  \over f_2(0.4,k_\perp)  }~;
\end{align}
in   { fact }, if a factorized ansatz, {\bf e.g.} 
$f_2(x,k_\perp) \sim f_{2,x}(x) f_{2, k_\perp}(k_\perp)$, { is used, }
 $r_k(x, k_\perp)$   { does} not depend on 
$k_\perp$ ~\cite{noi1,noi2,melo}. { The factorization ansatz is often used 
in experimental analyses for the proton target.}

{ In closing this section, we  { note} that  the
dPDFs { depend} on two momentum scales, corresponding to the mass of the states 
produced in the two 
parton-parton scattering in the DPS process,
{ which have not been 
explicitly shown.} } 

\subsection{
\label{IIA}An approximation in terms of one body quantities}

 An  { ansatz}  commonly used to  
describe the unknown dPDFs   { makes use of ffs and GPDs} (in the case of the 
proton
some experimental knowledge is available). Following the strategy of Refs. 
\cite{Diehl1,blok1,blok2}, we consider the correlator  (\ref{f2c1}) and 
insert 
a complete set of states  assuming that the pion is  
dominant. The formal expression for this approximated quantity, {  
  $f'_{2,A}(x_1,x_2, {\bf k_\perp} )$, } is:

\begin{align}
\nonumber
 & f'_{2,A}(x_1,x_2, {\bf k_\perp} )=
 {P^+ \over 4 } \int d^2 {\bf y_\perp} 
e^{-i 
{\bf 
y_\perp} \cdot {\bf k_\perp} }\int dy^- 
~
\\
&\times
 \int dz^-_1dz_2^- \int \dfrac{dP'^+d^2 {\bf P'_\perp}  
}{2(2\pi)^3 P'^+} {e^{-i x_1P^+ z^-_1-i 
x_2P^+ z^-_2} \over (2\pi)^2 }
\label{f2c1a}
\\
\nonumber
& \times
  \langle A, {\bf 0}| \mathcal{O}(0,z_1) |A',{\bf 
P'_\perp}\rangle \langle A', {\bf P'_\perp}|
\mathcal{O}(y,z_2)|A, {\bf 
 {\bf 0}} \rangle \Big|_{y^+ =z^+_1=z_2^+=0}^{{\bf z_{1\perp}= z_{2\perp}  }=0  
}~.
\end{align}
{  In this scenario, the approximation relies on the assumption 
$f'_2(x_1,x_2, {\bf k_\perp} ) \sim f'_{2,A}(x_1,x_2, {\bf k_\perp} ).$  }
At this point, using again  the strategy already 
discussed in the previous section, we find:

\begin{align}
 \label{f2ma1}
 f'_{2,A}(x_1,x_2,{\bf k_\perp}  ) &= H(x_1,{\bf k_\perp} )H(1-x_2,{\bf 
k_\perp} 
)~,
\end{align}
where $H(x, {\bf k_\perp})=H(x, \xi=0,{\bf k_{\perp}})$, is the pion GPD at zero 
skewness.
{  The integral over $x_2$ of Eqs. (\ref{f2m1}) and (\ref{f2ma1}) leads 
approximately to }

\begin{align}
 \label{f2ma}
 f_{2}(x,{\bf k_\perp}  ) \sim \int_0^1d x_2~f'_{2,A}(x,x_2,{\bf k_\perp} ) =  
H(x,{\bf k_\perp} )F({\bf k_\perp} 
)~,
\end{align}
where $F(k_\perp)$ { is} the standard {  pion } e.m. form factor.
{ The difference between $f_2(x,k_\perp)$ and $H(x, {\bf k_\perp})F({\bf 
k_\perp})$ addresses the presence of unknown parton correlations that can not 
be studied  by means of one-body distributions. In order to emphasize such 
effects, }
{ the  relation (\ref{f2ma}) will be { discussed } in the next section.}

The GPD for the pion \cite{universal} 
might be written also in terms of the wave function \cite{dihel_gpd,gpd},

\begin{align}
 H(x,\xi=0, \Delta_\perp^2) &= {1 \over 2} \sum_{h, \bar h} \int \dfrac{d^2 {\bf 
k_\perp} 
 }{16 \pi^3}
 \label{GPD}
 \\
 \nonumber
&\times \psi_{h, \bar h}\big(x ,  {\bf k_\perp }   \big)
\psi^{*}_{h, \bar h}\big(x,  {\bf k_\perp }  + {(1-x)}  
{\bf \Delta_\perp }   \big)~,
 \end{align}
\\
 { an } expression  well suited for model 
calculations { which} will be  { used} 
in the next section.

\subsection{\label{IIB}   The effective cross section }

A relevant 
 observable { for { DPS} proton studies}    is the so called 
effective 
cross section, $\sigma_{eff}$, { see e.g. Ref. \cite{fabbro}}. 
It is defined as the ratio of the product of two single parton 
scattering process 
cross sections to the DPS  with { the } same final states. 
{ It is extracted from data using model assumptions,}
and it can be { expressed}  in
terms of PDFs and dPDFs~\cite{plb}.
For proton-proton collisions, { this { quantity}  has been { also } studied} 
within the AdS/QCD soft-wall
model  \cite{Traini:2016jru}. { In }
 Refs. \cite{plb,Traini:2016jru} { it has been shown}   how a 
dependence of $\sigma_{eff}$ on the longitudinal momentum 
fractions of { the}  acting partons reflects the presence of non trivial  
double parton correlations. 
  In the present  { study }  
we {  use} { the}  
definition of $\sigma_{eff}$ { for a }  meson 
{ target}  
{ in order} to make { new } predictions.

The effective cross section 
for a DPS process, involving meson-meson collisions, generally depends on four 
variable $x_1,x_2$ and $x_1',x_2'$, i.e. the longitudinal momentum fractions of 
the partons  involved in the process. Nevertheless
in the  zero rapidity 
region, i.e. $x_1=x_1'$ and $x_2=x_2'$,  $\sigma_{eff}$ reads:

\begin{align}
 \label{sieff}
 \sigma_{eff}(x_1) = { \big(f_1(x_1)f_1(1-x_1)\big)^2   \over
 \int { d^2 {\bf k_\perp} \over  (2\pi)^2 } f_2(x_1, {\bf k_\perp} )^2  }~,
\end{align}
\\ { where }  $f_1(x)$ { is }  the {  single} PDF.
Furthermore, one can define { an average}
value  as 
follows:

\begin{align}
 \label{sieffa}
 \overline{ \sigma}_{eff}  = {1 \over \int { d^2 {\bf k_\perp} \over  (2\pi)^2 
} 
 F_{2}( {\bf k_\perp} )F_{2}(- {\bf k_\perp} )  }~,
\end{align}
where the effective form factor 

\begin{align}
\label{eff}
 F_{2}( {\bf k_\perp} ) = \int_0^1dx~f_2(x, {\bf k_\perp } )
\end{align}
\\
{ has been introduced  (see Refs. \cite{plb,ffe}).}
{ Equation} (\ref{sieffa})  { assumes }  
factorization between the $x$ 
and 
 $k_\perp$   { in} the dPDF.
In this factorized scenario,
one might notice 
that $\sigma_{DPS}$, i.e. the DPS cross section (see, e.g., Refs.
\cite{Paver:1982yp,noiW}), depends on  $1/\overline{\sigma}_{eff}$ \cite{noiW}. 
Thanks to this feature, the value of $1/\overline{\sigma}_{eff}$ 
provides a 
rough estimate of the   { magnitude} of $\sigma_{DPS}$.   
{ In our model calculation, } 
we 
provide  predictions for 
{ hypotetical experiments with mesons,} { as illustrated in the next 
section.}

\section{Calculation of the pion dPDF using AdS/QCD models}
\label{III}

{  In  { the present}   section we introduce and discuss  the LF 
wave 
function { then} used  to  { 
evaluate} the dPDFs.}{ In  { particular},  { we will make use 
of} the approach based on the { AdS/QCD} soft-wall model, { where}  
 } 
the pion w.f.  
reads \cite{Br1,Br2}:

\begin{align}
\label{pionwfori}
\psi_{\pi o} (x,{\bf k_{1 \perp}}  )= A_o {4\pi 
\over  \kappa_o \sqrt{x(1-x)} } 
e^{- { {\bf k^2_{1\perp} } + m_o^2 \over x(1-x) 2\kappa^2_o  }  }~,
\end{align}
where  $m_o=m_u \sim m_{\bar d}$, $x=x_1$, $x_2 = 1-x_1$ and 
${\bf k_{2\perp} } = - {\bf k_{1\perp}  } $.  The parameters of the model 
have been {  recently } fixed to
 reproduce  the Regge  behavior  of the mass spectrum of 
 mesons ~\cite{15sp,spin}.
 They are $\kappa_o = 0.523$ GeV and $m_o \sim 0.33$ GeV. 
 The constant $A_o$, is  fixed by the  
{normalization} condition  (\ref{norm}) and it is found to be $A_o = 
3.0498$. 
{  Several models of pion LF wave functions are available,
\cite{Br1,Br2,universal,spin},
 the most  { straightforward and therefore }  suitable
to show general properties of pion dPDFs  
{ is probably the first model proposed \cite{Br1,Br2}, where the dPDF is 
analytically expressed by:  }

\begin{align}
 \label{dpdfpiono}
 &f_2^{\pi O}(x, {\bf k_\perp} ) = A_o^2 e^{- { 4m_o^2+  {\bf k_\perp}^2 \over 
4 
\kappa_o^2 
x (1-x) }  }~.
\end{align}
 
 { In this paper, we calculate dPDF for $\pi^+$. }
 The distributions for $\pi^-$ and $\pi^0$ can be
obtained 
by isospin and charge conjugation. 
As one can see in the left 
panel of Fig.  \ref{dpdfpo}, 
{ as   happens in the proton case~\cite{noi2, noi1},}
the dPDF decreases as $k_\perp$ increases, 
and the  factorization in the $k_\perp$ and $x$  
 is not 
supported by the model as  can be observed in 
the right 
panel of Fig. \ref{dpdfpo}, where  the     { the ratio of}  Eq. (\ref{rk}) 
{ shows a clear $\mathbf{ k_\perp}$ 
dependence}.
{ We conclude the discussion of these results}
by reporting  the mean value of $\sigma_{eff}$ 
within the  model at the hadronic scale 
$\mu_0$,  $\overline{\sigma}_{eff}^{\pi} (\mu_0)= 41.69  $ mb. 
This value is larger than  the { corresponding} proton { 
value}  \cite{plb,blok1,blok2}, { a} 
 feature  
related to the geometrical properties of the targets 
{  (see Ref. \cite{ffe} for 
details). }

 \begin{figure*}[htb]
\includegraphics[scale=0.80]{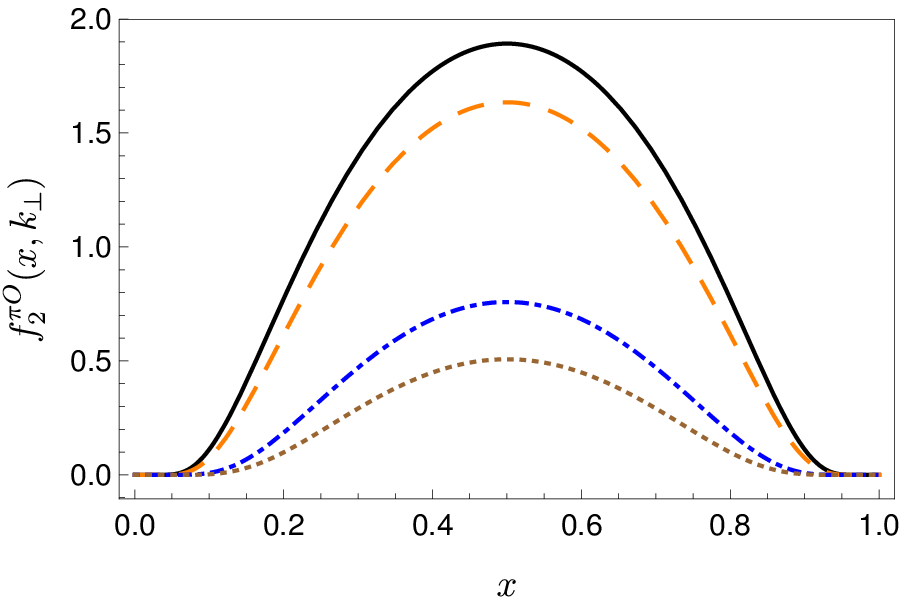}
\hskip 1cm \includegraphics[scale=0.80]{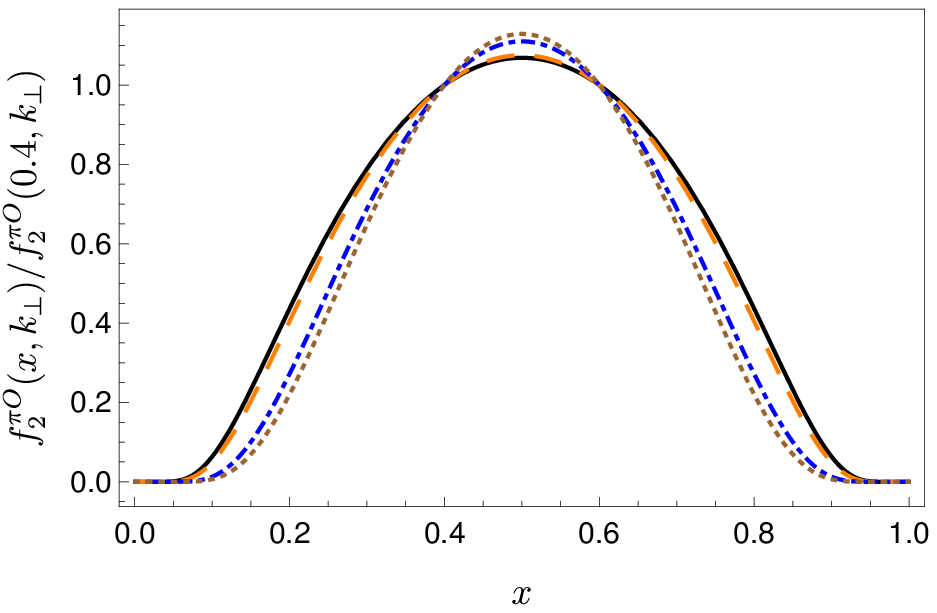}
\caption{\footnotesize  \textsl{Left panel: dPDF of the pion  { within} the 
AdS/QCD  model of Ref. 
\cite{Br1} {  (cfr Eq. (\ref{pionwfori}))}
   { at} 
different values of $k_\perp$. Full line $k_\perp=0$ GeV, dashed line 
$k_\perp=0.2$ GeV, dot-dashed line $k_\perp=0.5$ GeV and dotted line $k_\perp 
= 0.6$ GeV. Right panel: The ratio defined by Eq. (\ref{rk}) for the same 
parameters as in the left panel. }}
\label{dpdfpo}
\end{figure*}
 
\noindent
For completeness, we  report in Fig. \ref{H_o_v} the pion GPD 
evaluated  within the  model. As one can see, the pion 
{ GPD is} very 
similar to its dPDF. It is apparent that the expressions for the dPDF and
GPD, Eqs. (\ref{f2m},\ref{GPD}),
 in terms of the light-front pion wave function, are similar.
 However, in  the  dPDF, $k_\perp$ 
represents an intrinsic imbalance of the parton momentum between the 
initial and the final states keeping the same pion momentum in both states,
while in the GPDs, $\Delta_\perp=k_\perp$ represents the difference 
in momentum between 
the initial and final state of the pion.
Therefore, the dependence of the GPDs on the partonic momentum, i.e.  ${\bf 
k_{1,\perp} } \pm (1-x){\bf k_\perp}$ produces an asymmetry in the $x$ 
dependence, which is not  present in dPDF. Moreover, since in the GPDs the
momentum 
imbalance in the wave function is multiplied by the pre-factor $1-x<1$, at  
variance with the dPDF, the latter  goes to zero faster then the GPD.
{ 
Let us stress that such a similarity between dPDFs and GPDs 
holds only for the valence  { component and} at the hadronic 
scale, 
i.e. where only two { valence} particles
 { are} taken into account in the model.  If higher Fock 
states  { were} included 
in the LF wf representation of the pion, { other } non perturbative $x_1-x_2 
$ 
correlations { would} appear. 
Moreover if one considers the pQCD evolution of 
dPDFs, 
also perturbative $x_1-x_2$ correlations show up (see e.g. Ref. 
\cite{JHEP2016}).   { Analogously to the proton case, all} these 
non trivial dependence of dPDFs on $x_1$ and $x_2$ 
cannot be accessed via GPDs,   { a 
confirmation of the rich three-dimensional structure accessible via dPDFs}.   
 }

Finally we compare { the complete} $f_2^\pi (x, {\bf k_\perp})$ with its 
approximation   Eq. 
(\ref{f2ma}), i.e. $f_{2,A}^\pi (x, 
{\bf k_\perp})$.
If only the valence contribution  { were} considered, 
the approximation to the dPDF { would} become a product 
of a GPD and  { a} form factor, as seen in Eq. (\ref{f2ma}),  at 
variance with the proton case, where the dPDF is written as a product 
of two GPDs. In  Fig. \ref{App_o_v}, we 
{ compare} the dPDF  (\ref{f2m}) and its approximation 
(\ref{f2ma}) 
as a function of $x$ for three different values of $k_\perp$. 
As one can see, at the hadronic scale,  the pion dPDFs contains non trivial
information different from { that}  encoded 
in the GPDs and ffs, a feature also  {  
observed } in the proton case (see e.g. Refs. 
\cite{bag,noi2,noi1,JHEP2016,melo,noiW,ffe,seff}).

\begin{figure}
\begin{center}
\includegraphics[scale=1.]{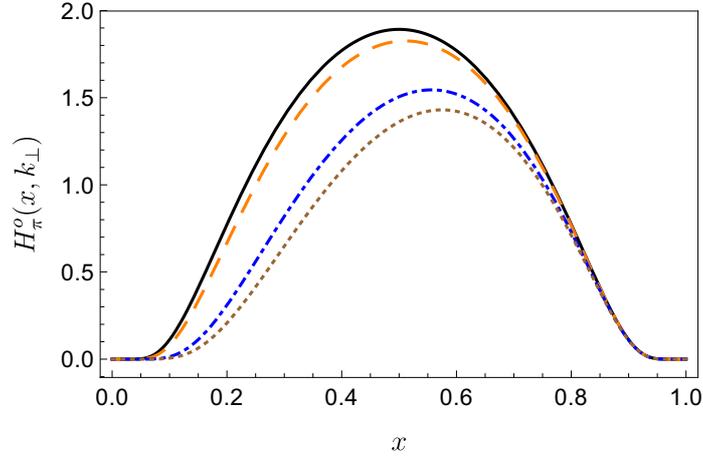}
\end{center}
\caption{\footnotesize  \textsl{The pion GPD defined in Eq. (\ref{GPD})
for  $k_\perp=0$ GeV (full line), $k_\perp=0.2$ GeV (dashed line), 
$k_\perp=0.5$ GeV (dot-dashed) line and $k_\perp=0.6$ GeV (dotted line).  }}
\label{H_o_v}
\end{figure}

\begin{figure}
\begin{center}
\includegraphics[scale=1.]{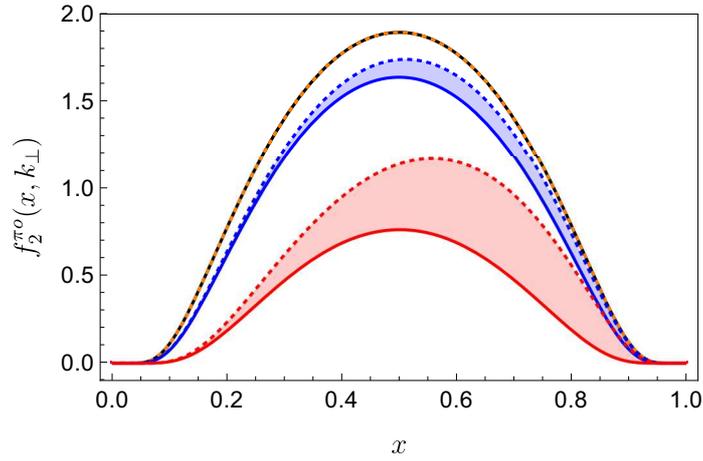}
\end{center}
\caption{\footnotesize  \textsl{The pion dPDF,
evaluated by means of its definition  { of}  Eq. (\ref{f2m}), 
is shown 
in full lines, and its 
approximation, defined by Eq. (\ref{f2ma}), is plotted in dotted lines, for
three values of $k_\perp$:
$k_\perp=0$ GeV, $k_\perp=0.2$ GeV and $k_\perp=0.5$ GeV. The quality of the 
approximation decreases as $k_\perp$ increases as shown by the 
bands  { emphasizing} for the difference between the exact 
calculation and 
the approximation. }}
\label{App_o_v}
\end{figure}

\section{\label{IV} Evolution} 
 The next step in our scheme is to 
calculate 
the perturbative evolution of the dPDFs from the { low momentum scale
of the model, the so called}
hadronic scale $\mu_0^2$, to the high scale
of the data $Q^2$. 
{As stated in the introduction,  dPDFs depend on two momentum scales.
For simplicity, as it has been in done in previous works, see e.g. Ref. 
\cite{evo13}, we assume here that 
two scales coincide. }
We follow here the same 
strategy developed in Refs. \cite{noi1,JHEP2016} 
adapted to the use of quark models to calculate the proton's dPDFs.
Historically, the evolution equations for dPDFs can be seen as a 
generalization of the usual DGLAP equations
{ (see the original papers  \cite{Kirschner:1979im,Shelest:1982dg}
and recent contributions in Refs.} 
\cite{Diehl1,blok1,evo13,evo2,evo3,evo4,evo5,evo6,evo7,evo8,evo10,evo11,evo12}
). 
This feature sets up the strategy which we are going to discuss next.

We start with the decomposition of the dPDF  at a generic scale $Q^2$: 
\begin{align}
F_{u \bar d} & =  F_{(u_V + u_{sea}) (\bar d_V + \bar d_{sea})} = F_{(u_V + 
\bar u) (\bar d_V + \bar d)} = \nonumber \\
& =  F_{u_V \bar d_V} + F_{u_V \bar d} + F_{\bar u \bar d_V} + F_{\bar u \bar 
d} 
\label{eq:Fubard}
\end{align}
where, at the hadronic scale \cite{GRS1998}, 
\be
u_V^{\pi^+} = \bar d_V^{\pi^+} = d_V^{\pi^-} = \bar u_V^{\pi^-} = {\it  v}^\pi 
\equiv u_V;
\label{eq:isospinsymm}
\ee
while at any scale  $q_{sea} = \bar 
q_{sea} = \bar q$, with $q=u,d,s$ for $N_f = 3$ { three active flavors}. 
It is convenient to use the symmetrized  form of dPDFs, 
$\bar F_{ab}= (F_{ab}+F_{ba} )/2 $ where 
$\bar F_{ab} \equiv \bar F_{a b}(x_1,x_2,{\bf k}_\perp, Q^2)$ { is} 
symmetric in 
$x_1$, $x_2$.

%

\subsection{Flavor decomposition   } 
In order to proceed with the evolution equations one has to construct 
from the $\bar{F}_{u \bar{d}}$ the Singlet and 
Non-Singlet  { components}

\begin{align}
\nonumber
& \Sigma = \sum_q q^+  = u_V + 2 \bar u + 
d_V + 2 \bar d + s + \bar s~\\
\nonumber
&T_3  =  u^+ - d^+  = 
u_V + 2 \bar u - d_V - 2 \bar 
d
\\
\nonumber
& T_8  =   u_V 
+ 
2 \bar u + d_V + 2 \bar d - 2 (s + \bar s)
\\
& V_i = q_i^-~,
\end{align}
where $q_i^\pm=q_i \pm \bar q_i$.

The evolution equations involve different equations  for the 
$Singlet-Singlet$ component 
($\Sigma 
\Sigma$),
$Non Singlet-Singlet$ components 
($T_8 \Sigma+\Sigma T_8$, $d_V \Sigma+\Sigma d_V$, $u_V 
\Sigma+\Sigma u_V$), and  $Non Singlet -Non Singlet$ contributions 
(constructed from $V_i$, $T_3$, $T_8$).

\subsection{$Mellin$-Moments and inversion}
The procedure follows by constructing  the 
$Mellin$-moments which allow to solve the evolution equations easily.
These quantities are

\begin{align}
\label{eq:Momijn1n2}
&   {1 \over 2} M_{(q_a q_b+q_b q_a)}^{n_1 n_2}(Q^2) = {  M_{q_b q_a}^{n_1 
n_2}(Q^2)+M_{q_a q_b}^{n_1 n_2}(Q^2)  \over 2} 
\\
\nonumber
& = \int_0^1 dx_1 \int_0^{1-x_1} 
dx_2 \, x_1^{n_1-1} x_2^{n_2-1} \,\bar F_{a b}(x_1,x_2; Q^2) .
\end{align}

{ At the hadronic scale $\mu_0^2$, all the combinations of dPDFs with 
$\Sigma,~T_8,~T_3$ and $V_i$ will contain valence partons only. As a result the 
remaining term will be $F_{u_V \bar d_V}$, and the non vanishing moments at 
the hadronic scale will assume the form  }

\begin{align}
\nonumber
M_{u_V \bar d_V}^{n_1 n_2}(\mu_0^2,{\bf k}_\perp) 
&= \int_0^1 dx_1 \int_0^{1-x_1} dx_2~ \delta(1- x_1 - x_2) 
\\
\label{mom}
&\times x_1^{n_1-1} x_2^{n_2-1}  f^{\pi 
O}_2(x_1,x_2,{\bf k}_\perp,\mu_0^2)  \\
\nonumber
&= \int_0^1 dx x^{n_1-1} (1-x)^{n_2-1}\,f^{\pi O}_2(x,{\bf k}_\perp,\mu_0^2).
\end{align}
They enter the moments of the combinations directly depending on 
$\Sigma,~T_3,~T_8$ and $V_i$, but each 
moment $M_{a b}^{n_1 n_2}(\mu_0^2)$, defined at the hadronic scale, will evolve 
according to its specific flavor symmetry \cite{JHEP2016}.
The moments are independent functions of the complex indices $n_1$, $n_2$ and 
 { the } inversion   { of Eq. (\ref{mom}) }  will produce 
dPDFs 
defined in the whole $ (x_1,x_2)$  domain with $x_1+x_2 \leq 1$. 

\subsection{Evolution of the dPDFs: results}

In Fig.\ref{fig:M22ubardpiOQ2}, we plot the second moment of the double
distribution $ x_1 x_2 \bar F_{u \bar d}(x_1, x_2, y, Q^2)$
 defined by
 \be
M_{u \bar d}^{2 2}(y,Q^2) = \int_0^1 dx_1 \int_0^{1-x_1} dx_2\,x_1 x_2 \bar
F_{u 
\bar d}(x_1, x_2, y, Q^2),
\label{eq:M22ubardpiOQ2}
\ee
where $y$ is the distance between the two 
correlated partons,  obtained Fourier transforming the ${\bf k}_\perp$ dependent 
distribution $f_2^{\pi_O}(x, {\bf k_\perp} )$ given in Eq. 
(\ref{dpdfpiono}). This quantity incorporates the evolution to large $Q^2$ of
the distribution  
$\bar F_{u \bar d}(x_1, x_2, y, Q^2)$  { starting from the initial scale  
$\mu_0^2=(0.523)^2$ GeV$^2$, already used  in calculation of pion PDF and 
unpolarized transverse momentum dependent PDF in Ref. \cite{pasquini1}}.

\begin{figure}
\centering\includegraphics[scale=0.4]
{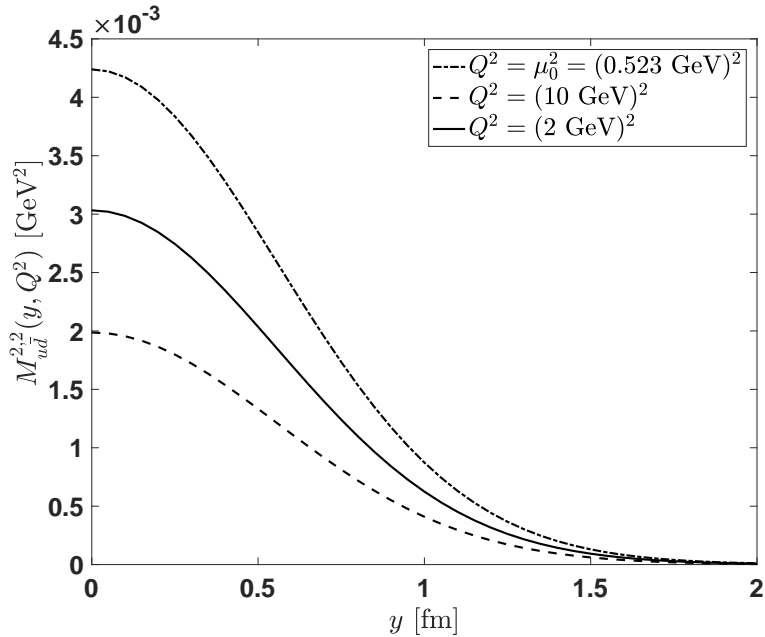}
\caption{\small   The second moment of { {dPDFs}},   
Eq.(\ref{eq:M22ubardpiOQ2}) as a
function 
of the distance $y$ and at different scales: the hadronic scale 
$\mu_0^2=(0.523 GeV)^2 $, $Q^2 = 4$ { GeV$^2$ } and 100 GeV$^2$.}
\label{fig:M22ubardpiOQ2}
\end{figure}

 At the hadronic scale since only two valence particles are present, the 
 support condition, preserved within the Light-Front approach, 
 forces the dPDF to exist only when $x_1+ x_2= 1$ . This is at variance with 
 the proton case where the existence of a third particle allows complete 
freedom for $x_1$ and $x_2$ as long as momentum is conserved $x_1+x_2<1$. The 
evolution procedure, described by  Eq. 
(\ref{eq:Momijn1n2}), where $n_1$ and $n_2$ are  independent complex 
parameters allows to obtain 
$F_{u \bar d}(x_1,x_2,y,Q^2)$ for all values of $x_1$ and $x_2$   { 
and}
$x_1+x_2<1$. The creation, in the evolution process, of sea and glue partons  
allows $x_1$ and $x_2$ to free themselves  from the valence condition.

\begin{figure}
\centering\includegraphics[scale=0.4]
{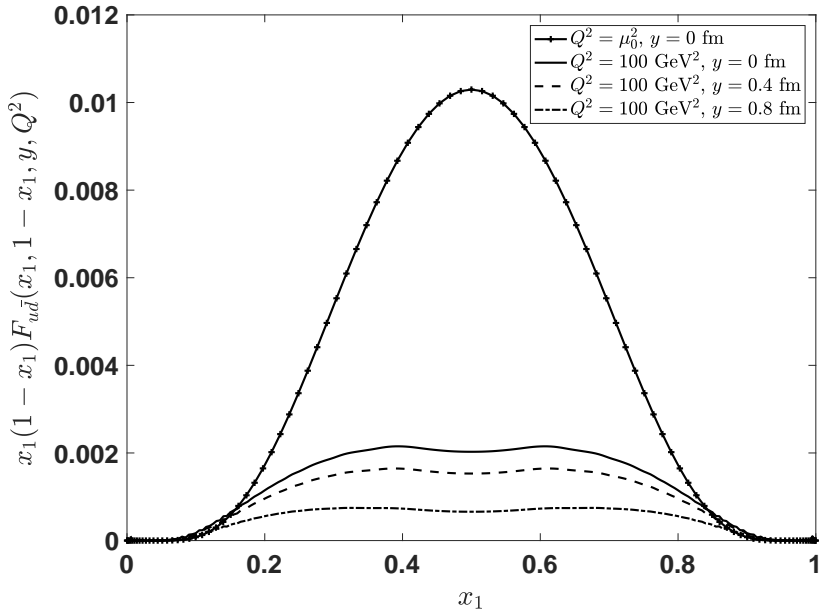}
\caption{\small   { The quantity } $x_1 (1-x_1) F_{u \bar 
d}(x_1, 1-x_1, y, Q^2=100~\mbox{GeV}^2)$ { is plotted} as a 
function of  
$x_1$ at different $y-$values. The input distribution at $\mu_0^2$ and 
$y=0$ fm is also shown. 
 }
\label{fig:x1mxFud_y_Q2}
\end{figure}

{ 
As we mentioned in the Introduction, preliminary results
for quantities related to moments of the pion dPDFs have been
recently reported { within a lattice QCD approach} 
\cite{Zimmermann:2017ctb}.  A comparison of results obtained in model 
calculations 
with lattice data
 { would open 
interesting new perspectives}. 
}

A first important effect of the evolution procedure  can be 
seen in  Fig.\ref{fig:x1mxFud_y_Q2},
 where the double 
distribution $x_1 x_2 \bar F_{u \bar d}(x_1,x_2,y,Q^2 = 100\,{\rm GeV}^2)$ is 
shown in the domain ($x_1$, $x_2 = 1-x_1$) as a function of $x_1$ and for 
different values of $y$. The comparison with the same distribution at the 
hadronic scale $\mu_0^2$ and $y=0$ clearly emphasizes the effects of the 
evolution. The evolution from 
$\mu_0^2$ to $Q^2=100$ GeV$^2$ produces a reduction of the distribution, 
 a behavior  physically interpretable as the creation of new partonic species 
carrying momentum, in particular gluon distributions. Recall that the latter
are zero at the 
hadronic scale for the models considered.
In Fig.\ref{fig:xxFuVg_y_Q2}  the double distribution $x_1 x_2 
\bar F_{u_V g}(x_1,x_2,y,Q^2)$ is plotted. The upper panel 
shows the dependence of the distribution on the scale $Q^2$, while the lower 
panel illustrates its dependence on the parton distance $y$.

\begin{figure}
\centering\includegraphics[scale=0.4]
{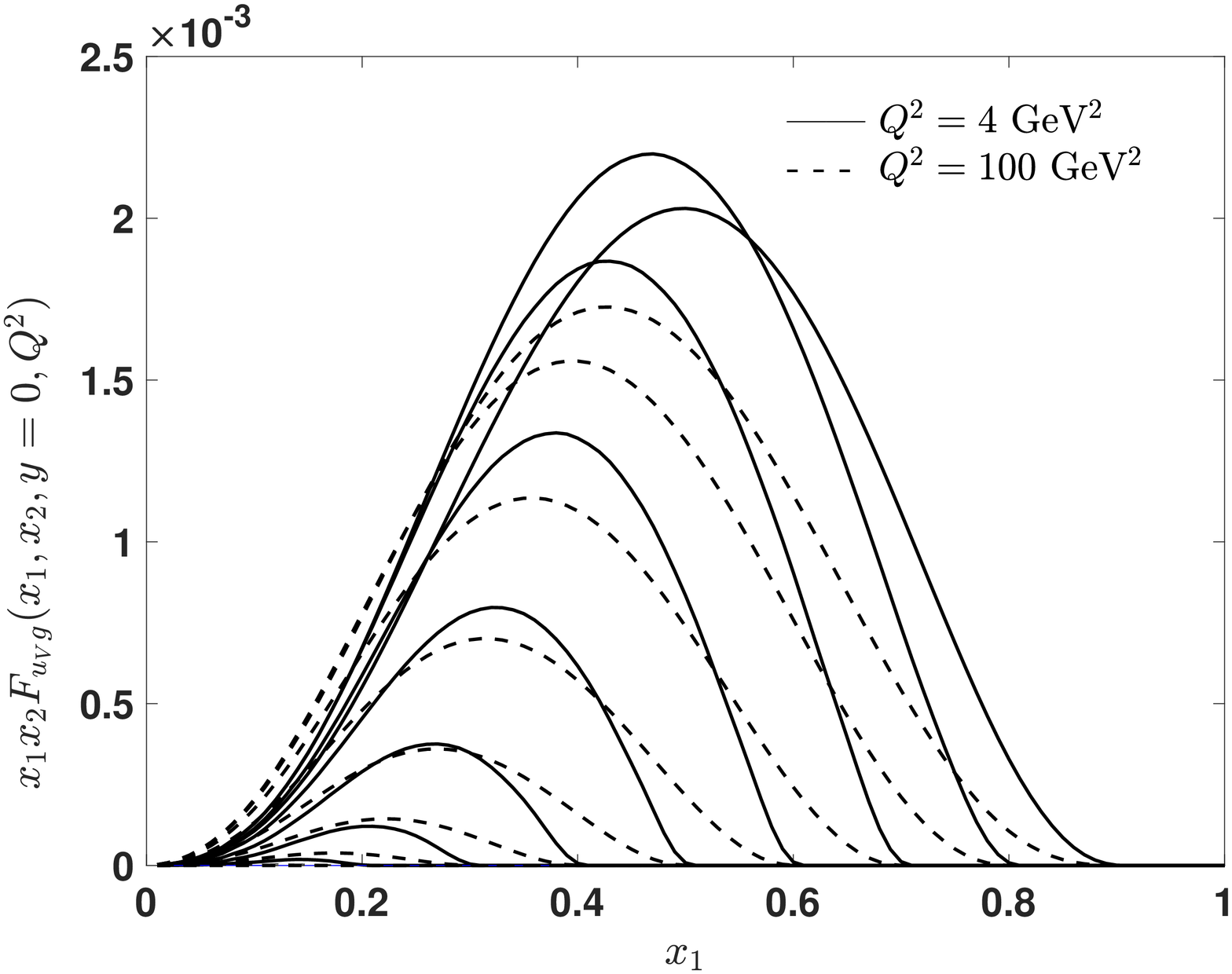}
\centering\includegraphics[scale=0.4]
{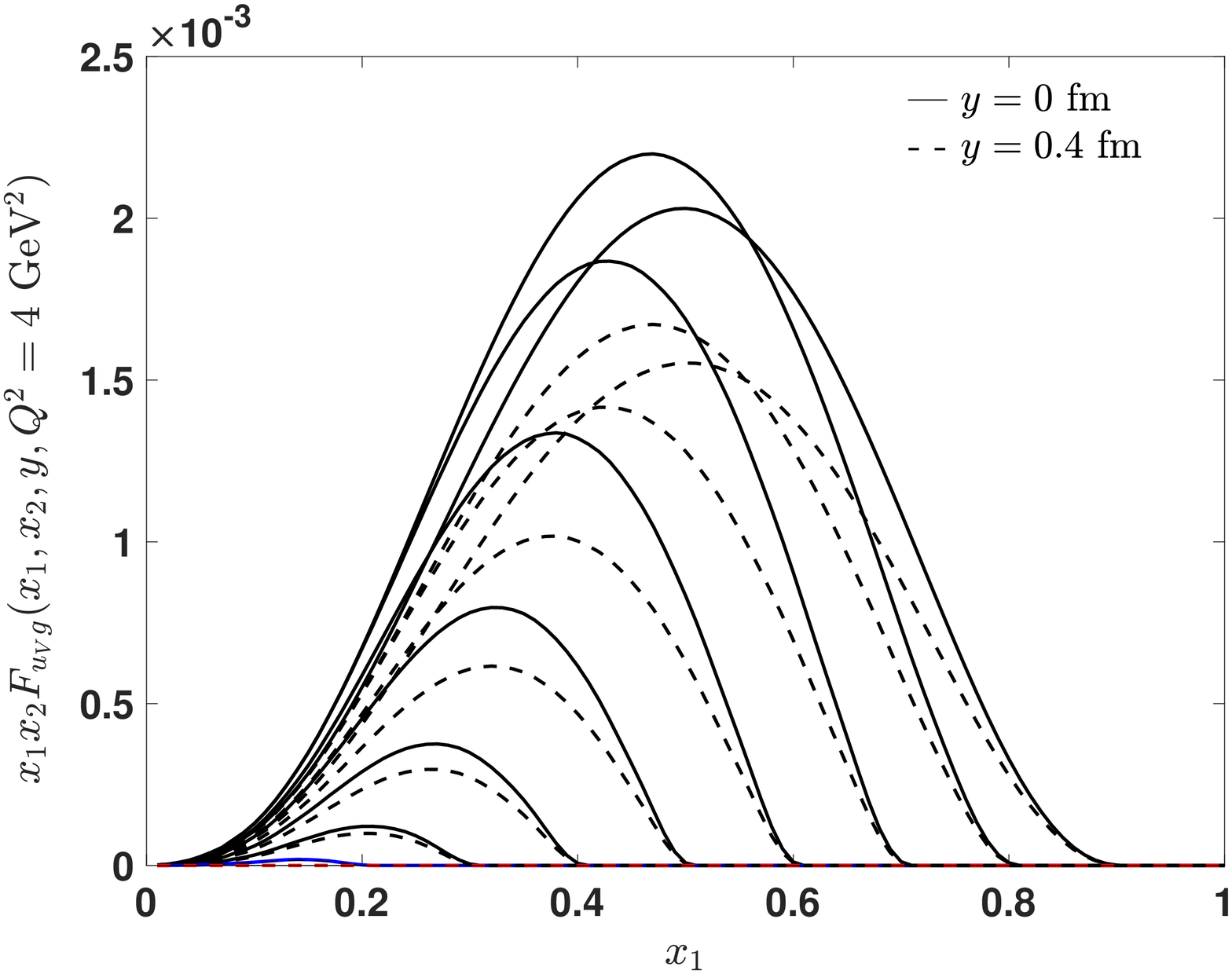}
\caption{\small  { The quantity} $x_1 x_2 F_{u_V g}(x_1, x_2, 
y, Q^2)$ { is plotted} as a 
function of  $x_1$ for $x_2 = 0.1,0.2,0.3,0.4,0.5,0.6,0.7,0.8,0.9$. In the 
upper panel the full 
lines represent the results at $Q^2=4$ GeV$^2$ and the dashed lines those ones 
at $Q^2=100$ GeV$^2$ for the same value of the parton distance $y=0$ fm. The 
lower panel compares the distributions  for 
different distances: $y=0$ fm (full lines) and $y=0.4$ fm (dashed lines) { 
and $Q^2=4$ GeV$^2$.} }
\label{fig:xxFuVg_y_Q2}
\end{figure}

A large part of the valence parton momentum is transferred  to the gluons 
which  increases dramatically at low-$x$, while the relevance of the valence
partons 
decreases.


%

\section{\label{VI} Conclusions }
{{  
{ Double parton distribution functions may } 
represent a novel tool to access the three 
dimensional structure of hadrons. It is therefore natural to study the dPDFs 
of 
the pions, specially now that the first estimates of quantities related to the 
dPDF of pion have been reported by lattice studies \cite{Zimmermann:2017ctb}. 
{ We have used here}
 a Light Front formalism, { for which} 
the wave function of the system is required.
The AdS/QCD correspondence  { 
has generated our} 
  LF  wave function.
  { Once the formalism has been set up we have calculated several quantities}
{ Among them,} we have obtained 
  the mean effective cross section 
for pion-pion scattering at the hadronic 
scale, $\bar \sigma_{eff}^\pi (\mu_0)= 41$ mb, which turns out to be 
larger than { the same cross section, evaluated with a similar approach,}
in the proton case. 
This quantity is very much  independent
{ on QCD evolution} and 
provides us with an estimate    
of the magnitude of DPS \cite{plb,Traini:2016jru}.

In the adopted AdS/QCD model, 
dPDFs turn out { to be} analytical.
{ It has been found that an approximation in terms of generalized
parton distributions, proposed in several approaches, is not reliable,
as it happens also}
in the proton case \cite{JHEP2016}. 
{ Analogously, our calculations show that dPDFs do not factorize into  
$x_{1,2}-$ and ${ k_\perp}-$ dependent terms. { These facts expose the 
presence of unknown double parton correlations in the pion not accessible from 
one-body distributions. }  }

We have performed the evolution to high $Q^2$ using the conventional 
formalism,  subject in this case, { at the model
momentum scale where only two valence constituents
with momentum fractions $x_1$ and $x_2$ are present,} 
to the $x_1+x_2=1$ restriction. 
{ Expected results are obtained. For example, 
the second moment decreases}
as $Q^2$ increases,  signalling the opening of 
new dPDFs associated with sea and gluons.  { A good example has been shown 
in Fig. \ref{fig:xxFuVg_y_Q2}, where the dPDF due to the correlation of valence 
and gluons is shown for different values of the parton distance and $Q^2$. At 
the hadronic scale such a distribution vanishes because no  gluons are 
included. At higher scale $Q^2$ the radiative production of gluons from the 
valence system makes  $F_{u_Vg}>0$.  } 
The dPDFs show, { both at the model scale and at a high momentum scale,} 
also a strong dependence on the partonic distance,
decreasing in magnitude as the distance increases.
 
{ While,} at present, 
experiments designed to measure dPDFs of the pion  { cannot be imagined}, 
 lattice calculations have started to approach 
this problem { and will be likely able, in the near future, to distinguish
between predictions of different
models of the pion structure, such as the one presented here, { opening new 
perspectives}}.
}}

\section*{Acknowledgements}
This work was supported in part by the Mineco under contract 
FPA2013-47443-C2-1-P and SEV-2014-0398.
{ M.T. and V.V. thank the University of Perugia and INFN, Perugia section,
for warm hospitality and support. S.S. thanks the University of Valencia 
and the IFIC for warm hospitality and support.}


\begin{thebibliography}{90}


\bibitem{Aad:2013bjm}
  G.~Aad {\it et al.} [ATLAS Collaboration],
  New J.\ Phys.\  {\bf 15} (2013) 033038
  
\bibitem{Paver:1982yp}
  N.~Paver and D.~Treleani,
  Nuovo Cim.\ A {\bf 70} (1982) 215.
  
\bibitem{Diehl1}
M.~Diehl, D.~Ostermeier and A.~Schafer,
  JHEP {\bf 03}, 089 (2012)

\bibitem{Guidal:2013rya}
  M.~Guidal, H.~Moutarde and M.~Vanderhaeghen,
  Rept.\ Prog.\ Phys.\  {\bf 76} (2013) 066202
  
  
\bibitem{Dupre:2016mai}
  R.~Dupr\'e, M.~Guidal and M.~Vanderhaeghen,
  Phys.\ Rev.\ D {\bf 95} (2017) no.1,  011501
 
\bibitem{bag}
  H.~M.~Chang, A.~V.~Manohar and W.~J.~Waalewijn,
  Phys.\ Rev.\ D {\bf 87} (2013) no.3,  034009.

\bibitem{noi2}
  M.~Rinaldi, S.~Scopetta and V.~Vento,
  Phys.\ Rev.\ D {\bf 87}, 114021 (2013)



 \bibitem{noi1} 
  M.~Rinaldi, S.~Scopetta, M.~Traini and V.~Vento,
  JHEP {\bf 12}, 028 (2014)



 
   \bibitem{plb} 
  M.~Rinaldi, S.~Scopetta, M.~Traini and V.~Vento,
  Phys.\ Lett.\ B {\bf 752}, 40 (2016).
  
   \bibitem{JHEP2016} 
  M.~Rinaldi, S.~Scopetta, M.~C.~Traini and V.~Vento,
  JHEP {\bf 16}, 063 (2016).

\bibitem{Traini:2016jru}
  M.~Traini, M.~Rinaldi, S.~Scopetta and V.~Vento,
  Phys.\ Lett.\ B {\bf 768} (2017) 270


\bibitem{Kirschner:1979im}
  R.~Kirschner,
  Phys.\ Lett.\  {\bf 84B} (1979) 266.

\bibitem{Shelest:1982dg}
  V.~P.~Shelest, A.~M.~Snigirev and G.~M.~Zinovev,
  Phys.\ Lett.\  {\bf 113B} (1982) 325.


\bibitem{Maldacena:1997re}
  J.~M.~Maldacena,
  Int.\ J.\ Theor.\ Phys.\  {\bf 38} (1999) 1113
   [Adv.\ Theor.\ Math.\ Phys.\  {\bf 2} (1998) 231]
  
\bibitem{Witten:1998qj}
  E.~Witten,
  Adv.\ Theor.\ Math.\ Phys.\  {\bf 2} (1998) 253


\bibitem{Polchinski:2000uf}
  J.~Polchinski and M.~J.~Strassler,
  hep-th/0003136.



\bibitem{Brodsky:2003px}
  S.~J.~Brodsky and G.~F.~de Teramond,
  Phys.\ Lett.\ B {\bf 582} (2004) 211
  
\bibitem{Erlich:2005qh}
  J.~Erlich, E.~Katz, D.~T.~Son and M.~A.~Stephanov,
  Phys.\ Rev.\ Lett.\  {\bf 95} (2005) 261602

\bibitem{DaRold:2005mxj}
  L.~Da Rold and A.~Pomarol,
  Nucl.\ Phys.\ B {\bf 721} (2005) 79




   \bibitem{Br1}
  S.~J.~Brodsky and G.~F.~de Teramond,
  Phys.\ Rev.\ D {\bf 77}, 056007 (2008).


      \bibitem{Br2}
  S.~J.~Brodsky and G.~F.~de Teramond,
  Subnucl.\ Ser.\  {\bf 45}, 139 (2009).
 
 \bibitem{rho}
  J.~R.~Forshaw and R.~Sandapen,
  Phys.\ Rev.\ Lett.\  {\bf 109}, 081601 (2012)

  
  { 
 \bibitem{universal} 
  G.~F.~de Teramond {\it et al.} [HLFHS Collaboration],
  Phys.\ Rev.\ Lett.\  {\bf 120}, no. 18, 182001 (2018)
  
  
  }
  
  
  
  
  

\bibitem{Traini:2016jko}
  M.~C.~Traini,
  Eur.\ Phys.\ J.\ C {\bf 77} (2017) no.4,  246


  \bibitem{rinaldiGPD}
  M.~Rinaldi,
  Phys.\ Lett.\ B {\bf 771}, 563 (2017)
  

  
  
  

 \bibitem{15sp}
  R.~Swarnkar and D.~Chakrabarti,
  Phys.\ Rev.\ D {\bf 92}, no. 7, 074023 (2015)


\bibitem{pasquini1}
  A.~Bacchetta, S.~Cotogno and B.~Pasquini,
  Phys.\ Lett.\ B {\bf 771}, 546 (2017).
  

{
\bibitem{Zimmermann:2017ctb} 
  C.~Zimmermann [RQCD Collaboration],
  PoS LATTICE {\bf 2016}, 152 (2016)
}


\bibitem{kase}
  T.~Kasemets and A.~Mukherjee,
  Phys.\ Rev.\ D {\bf 94}, no. 7, 074029 (2016)




\bibitem{48j1}
  S.~J.~Brodsky, H.~C.~Pauli and S.~S.~Pinsky,
  Phys.\ Rept.\  {\bf 301}, 299 (1998).
  
  \bibitem{24p}
  S.~J.~Brodsky, M.~Diehl and D.~S.~Hwang,
  Nucl.\ Phys.\ B {\bf 596}, 99 (2001).

 

  \bibitem{melo}
  M.~Rinaldi and F.~A.~Ceccopieri,
  Phys.\ Rev.\ D {\bf 95}, no. 3, 034040 (2017).


\bibitem{blok1}
  B.~Blok, Y.~Dokshitser, L.~Frankfurt and M.~Strikman,
  Eur.\ Phys.\ J.\ C {\bf 72}, 1963 (2012).

\bibitem{blok2}
  B.~Blok, Y.~Dokshitzer, L.~Frankfurt and M.~Strikman,
  Eur.\ Phys.\ J.\ C {\bf 74}, 2926 (2014).



 \bibitem{dihel_gpd}
  M.~Diehl,
  Phys.\ Rept.\  {\bf 388}, 41 (2003).

  
  
    
  \bibitem{gpd}
  C.~Mezrag, L.~Chang, H.~Moutarde, C.~D.~Roberts, J.~Rodr\'{\i}guez-Quintero, 
F.~Sabati\'e and S.~M.~Schmidt,
  Phys.\ Lett.\ B {\bf 741}, 190 (2015)
  
 
 \bibitem{fabbro}
  A.~Del Fabbro and D.~Treleani,
  Phys.\ Rev.\ D {\bf 63}, 057901 (2001)
  
  

  \bibitem{ffe}
  M.~Rinaldi and F.~A.~Ceccopieri,
  Phys.\ Rev.\ D {\bf 97}, no. 7, 071501 (2018)

  
  
  
  
    \bibitem{noiW}
  F.~A.~Ceccopieri, M.~Rinaldi and S.~Scopetta,
  Phys.\ Rev.\ D {\bf 95}, no. 11, 114030 (2017)
  
  
  \bibitem{spin}
  M.~Ahmady, F.~Chishtie and R.~Sandapen,
  Phys.\ Rev.\ D {\bf 95}, no. 7, 074008 (2017)
  


  
     \bibitem{seff}
  T.~Kasemets and S.~Scopetta,
  arXiv:1712.02884 [hep-ph].
  
   \bibitem{evo13}
  A.~M.~Snigirev, N.~A.~Snigireva and G.~M.~Zinovjev,
  Phys.\ Rev.\ D {\bf 90}, no. 1, 014015 (2014)
     
  

  \bibitem{evo2}
  A.~V.~Manohar and W.~J.~Waalewijn,
  Phys.\ Rev.\ D {\bf 85}, 114009 (2012)
  
  
  
 \bibitem{evo3} 
  E.~Cattaruzza, A.~Del Fabbro and D.~Treleani,
  Phys.\ Rev.\ D {\bf 72}, 034022 (2005)
  
 
 \bibitem{evo4}
  M.~Diehl and A.~Schafer,
  Phys.\ Lett.\ B {\bf 698}, 389 (2011)
  
  
 
 \bibitem{evo5}
  M.~G.~Ryskin and A.~M.~Snigirev,
  Phys.\ Rev.\ D {\bf 83}, 114047 (2011)
  
 
 \bibitem{evo6}
  J.~R.~Gaunt and W.~J.~Stirling,
  JHEP {\bf 1106}, 048 (2011)
  
 
 \bibitem{evo7}
  F.~A.~Ceccopieri,
  Phys.\ Lett.\ B {\bf 697}, 482 (2011)
  
 
 \bibitem{evo8}
  J.~R.~Gaunt,
  JHEP {\bf 1301}, 042 (2013)
  
  
   \bibitem{evo10}
  A.~V.~Manohar and W.~J.~Waalewijn,
  Phys.\ Lett.\ B {\bf 713}, 196 (2012)
  
  
  
  
  
  
\bibitem{evo11}
  W.~Broniowski and E.~Ruiz Arriola,
  Few Body Syst.\  {\bf 55}, 381 (2014)
 
 
 \bibitem{evo12}
  M.~Diehl, T.~Kasemets and S.~Keane,
  JHEP {\bf 1405}, 118 (2014)
 
 

  
  
  
  
  
    \bibitem{GRS1998}
  M.~Gluck, E.~Reya and M.~Stratmann,
  Eur.\ Phys.\ J.\ C {\bf 2}, 159 (1998).

  
 
  
  
  
  




 

  
 

    
 

  
  
 


  
 
    

  
  
  
  
  
   
  
 
  
  
    
  
  
  
  

  
  
    
   
  

  
  


  
  
  

  
  

  

  
 

  
 
 
  
   
\end{thebibliography}
\end{document}